# Renormalization of QED near Decoupling Temperature

Samina S. Masood

Physics Department, University of Houston Clear Lake, Houston, TX 77058
E-Mail: masood@uhcl.edu

## Abstract

We study the effective parameters of QED near decoupling temperatures and show that the QED perturbative series is convergent, at temperatures below the decoupling temperature. The renormalization constant of QED acquires different values if a system cools down from a hotter system to the electron mass temperature or heats up from a cooler system to the same temperature. At T = m, the first order contribution to the electron selfmass, δm/m is 0.0076 for a heating system and 0.0115 for a cooling system and the difference between two values is equal to 1/3 of the low temperature value and 1/2 of the high temperature value around T~m. This difference is a measure of hot fermion background at high temperatures. With the increase in release of more fermions at hotter temperatures, the fermion background contribution dominates and weak interactions have to be incorporated to understand the background effects.

## Introduction

Renormalization techniques of perturbation theory are used to calculate temperature dependence of renormalization constants of QED (quantum electrodynamics) at finite temperatures [1-11]. The values of electron mass, charge and wavefunction, at a given temperature, represent the effective parameters of QED at those temperatures. The magnetic moment of electrons, dynamically generated mass of photons and QED coupling constants are estimated as functions of temperature. However, thermal contributions to electric permittivity, magnetic permeability and dielectric constant of the medium are derived from the vacuum polarization. Some of the important parameters of QED plasma such as Debye shielding length, plasma frequency and the phase transitions can be obtained from the properties of the medium itself.

In this paper, we re-examine the analytical results of temperature dependent renormalization constants and prove that QED can safely be renormalized at finite temperatures, using the perturbation theory in a vacuum, below the neutrino decoupling temperature. We use the renormalization scheme of QED in real-time formalism [1-8] to calculate the electron mass, wavefunction and charge of electron as renormalization constants of QED [9-11]. It is now well-known that the existing first order calculations of the renormalization constants, in the real-time formalism, give the quadratic dependence of QED parameters on temperature T, expressed in units of electron mass m. Renormalization constants of QED, using the perturbation theory, give effective parameters of QED in a hot and dense medium and are very useful to understand the physics of the universe. However, the renormalization scheme is fully reliable below the decoupling temperature only. It is explicitly checked that in the existing scheme of calculations, the theory remains renormalizable at T ≤ 4m ~ 2MeV. However, the perturbative corrections will exceed the original values of QED parameters at higher temperatures and hard thermal loops have to be dealt with, using already developed methods [12-14]. However, below the neutrino decoupling temperature, the real part of the propagators is enough to describe the perturbative behavior of the system and doubling of the field is not required.

We discuss here the distinct behavior of QED below the neutrino decoupling temperatures. At these temperatures, QED coupling starts to play its role in modifying QED parameters for nucleosynthesis. With the help of these effective parameters of QED, the abundance of helium in the early universe can be estimated [12] precisely at a given temperature. The temperature dependent QED corrections to the nucleosynthesis parameters improve the results of the standard big bang model of cosmology [13-16] and it helps to relate the observational data, e.g.; WMAP (Wilkinson Microwave Anisotropy Probe [15] with the big bang theory. The same techniques can even be used to calculate perturbative effects in QCD [23] and electroweak processes [17] at low temperatures and estimate their contributions at high temperatures also. We assume that the change in the properties of neutrinos [18-22] does not affect the decoupling temperature significantly.

Without giving the calculational details, we have to briefly overview the existing form of QED renormalization constants in real-time formalism. The Feynman rules of vacuum theory are used with the statistically corrected propagators given as

$$D_\beta(k) = \frac{i}{k^2 - m^2 + i\varepsilon} + \frac{2\pi}{e^{\beta E_k} - 1}\delta(k^2 - m^2), \qquad (1)$$

for bosons, and

$$S_F(p) = \frac{i}{\not{p} - m + i\varepsilon} - \frac{2\pi(\not{p} + m)}{e^{\beta E_p} + 1}\delta(p^2 - m^2), \qquad (2)$$

for fermions in a hot medium. Since the temperature corrections appear as additive contributions to the fermion and boson propagators in Eqs (1) and (2), temperature dependent terms can be handled independently of vacuum terms at the one loop level. We restrict ourselves up to the two-loop level to show explicitly that second order thermal corrections are finite and smaller than the first order corrections [9], below the neutrino decoupling.

In the next section, we give the calculations of thermal corrections to the mass renormalization constant δm/m of QED and the physical mass of electrons at finite temperatures. Section 3 is devoted to the calculations of electron wavefunction at finite temperature. Section 4 gives the calculations of the electron charge and the QED coupling constant up to the two loop level, whereas in section 5, the last section is devoted to the discussions of the results of all the calculations to explicitly prove the renormalizeability of the theory up to the two loop level, below the decoupling temperature.

# Selfmass of Electron

The renormalized mass of electrons $m_R$ can be represented as a physical mass $m_{phys}$ of electrons and is defined in a hot and dense medium as,

$$m_R \equiv m_{phys} = m + \delta m(T=0) + \delta m(T). \qquad (3)$$

where m is the electron mass at zero temperature and $\delta m(T=0)$ represents the radiative corrections from a vacuum and $\delta m(T)$ are the contributions from the thermal background at nonzero temperature T. The physical mass of electrons can then be represented as a perturbative series in α and can be written as:

$$m_{phys} \cong m + \delta m^{(1)} + \delta m^{(2)} + \ldots \qquad (4)$$

where $\delta m^1$ and $\delta m^2$ are the shifts in electron mass in the first and second order in α, respectively. The physical mass is deduced by locating the pole of the fermion propagator $\frac{i(\not{p}+m)}{p^2-m^2+i\varepsilon}$ in thermal background. For this purpose, we sum over all of the same order diagrams. Renormalization is established by demonstrating the order-by-order cancellation of singularities. All the finite terms from the same order in α is combined together to evaluate the same order contribution to the physical mass given in eq (4). The physical mass in thermal backgrounds, up to order $\alpha^2$ [6,7], is calculated using the renormalization techniques of QED. Selfmass of electron is expressed as:

$$\Sigma(p) = A(p)E\gamma_0 - B(p)\vec{p}.\vec{\gamma} - C(p), \qquad (5)$$

where A(p), B(p), and C(p) are the relevant coefficients that are functions of electron momentum only. We take the inverse of the propagator with momentum and mass terms separated as:

$$S^{-1}(p) = (1-A)E\gamma^o - (1-B)p.\gamma - (m-C). \qquad (6)$$

The temperature-dependent radiative corrections to the electron mass up to the first order in α are obtained from the temperature dependent propagator as

$$m_{phys}^2 \equiv E^2 - |\mathbf{p}|^2 = m^2\left[1 - \frac{6\alpha}{\pi}b(m\beta)\right] + \frac{4\alpha}{\pi}\left[mT\,a(m\beta) + \frac{2}{3}\alpha\pi T^2 - \frac{6}{\pi^2}c(m\beta)\right]. \qquad (7)$$

giving

$$\frac{\delta m}{m} \simeq \frac{1}{2m^2}\left(m_{phys}^2 - m^2\right)$$

$$\simeq \frac{\alpha\pi T^2}{3m^2}\left[1 - \frac{6}{\pi^2}c(m\beta)\right] + \frac{2\alpha}{\pi}\frac{T}{m}a(m\beta) - \frac{3\alpha}{\pi}b(m\beta). \qquad (8)$$

where δm/m is the relative shift in electron mass due to finite temperatures which was determined in Ref. [1,9] with

$$a(m\beta) = \ln(1 + e^{-m\beta}), \qquad (9a)$$

$$b(m\beta) = \sum_{n=1}^{\infty}(-1)^n \text{Ei}(-nm\beta), \qquad (9b)$$

$$c(m\beta) = \sum_{n=1}^{\infty}(-1)^n \frac{e^{-nm\beta}}{n^2}, \qquad (9c)$$

The convergence of eq.(4) can be ensured for T<2MeV as δ m\m is always smaller than unity within this limit. This scheme of calculations will not work for higher temperatures and the first order corrections may exceed the original values of QED parameters, after 5 MeV. At the low temperatures T< m, the functions a(mβ ), b(mβ ), and c(mβ) fall off in powers of $e^{-m\beta}$ in comparison with $(T/m)^2$ and can be neglected in the low temperature limit, giving,

$$\frac{\delta m}{m} \xrightarrow{T<m} \frac{\alpha\pi T^2}{3m^2}. \qquad (10)$$

In the high-temperature limit, a(mβ ) and b(mβ ) are vanishingly small and the total fermion contribution comes from c(mβ )⟶ - $\pi^2$/12, yielding

$$\frac{\delta m}{m} \xrightarrow{T>m} \frac{\alpha\pi T^2}{2m^2}. \qquad (11)$$

The above equations give δm\m=7.647×$10^{-3}$($T^2$\$m^2$) for the low temperature and δm\m =1.147× $10^{-2}$ $T^2$\$m^2$ for the high temperature, showing that the rate of change of mass δm\m is larger at T>m as compared to T<m. Subtracting eq.(10) from (11), the change in δm\m between low and high temperature ranges can be written as

$$\Delta(\frac{\delta m}{m}) = \pm\frac{\alpha\pi T^2}{6m^2} = \pm 3.8 \times 10^{-3}\frac{T^2}{m^2} \qquad (12)$$

showing that the Δ(δm\m) = 3.8× $10^{-3}$ at T=m. δm\m = 0.0076 for a heating system and δm\m = 0.0115 for a cooling system and the difference between two values, Δ(δm\m) = 0.0038 such that Δ(δm\m) is equal to 1/3 of the low temperature value and 1/2 of the high temperature value at T=m. This difference is due to the photon background contributions at low temperatures and additional hot fermionic background at high temperatures. Therefore, the absence of hot fermion background contributes to a 50% decrease in selfmass as compared to cooling universe corresponding T value. The high T behavior will give 33% more selfmass as compared to the low T behavior. Since Δ(δm\m) quadratically grows with temperature, the fermion background contribution dominates over the hot boson background after the nucleosynthesis.

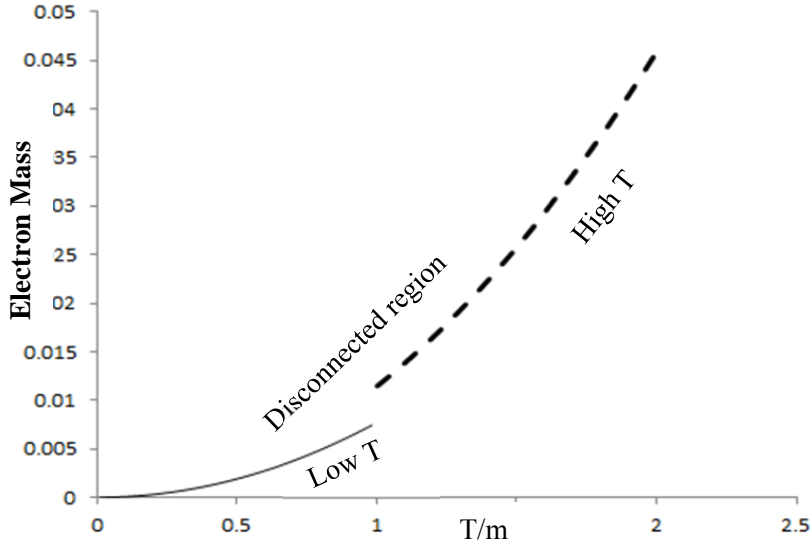

*Figure 1: A graph of selfmass of electron near T=m for a cooling system from higher temperature and a heating system from a lower temperature. Both High T and Low T curves have different slopes and neither value coincides.*

The temperature dependence of QED parameters is a little more complicated and significant during nucleosynthesis because of the change in matter composition during nucleosynthesis. Therefore eqs.(7) and (8) are required for the T~m region and help to estimate the change in QED statistical behavior due to the change in composition. This difference becomes more visible when we plot eqs. (12) and (13) of selfmass at low temperature and high temperature showing that both values start to give a disconnected region near T~m; i.e., the nucleosynthesis temperature. Figure 1 shows that the slope of both graphs (corresponding to eqs (10) and (11)) never meet at the common point T=m. The disconnected region around T~ m indicates a change in thermal properties for a heating and a cooling system. In a heating system, fermions start to produce around T~m whereas, in a cooling system fermions are eliminating around these temperatures. Gap between two curves around T~m is a measure of background fermion contribution.

That modification in the electron mass behavior in the range T~m is estimated by eq. (10). It is also clear from eq. (11) that after around 5MeV, the temperature dependence correction term ($\delta m \backslash m$) approaches to unity or even bigger for higher temperatures, even at the one-loop level. Higher order corrections [7,8] will also grow rapidly at high temperatures, giving

$$m_{phys} = m\left[1 + \frac{\delta m}{m} + \frac{1}{2}(\frac{\delta m}{m})^2 + \ldots\right] \approx m\exp(\frac{\delta m}{m}) \qquad (13)$$

A change occurs around T ~ m and is clearly related to nucleosynthesis where, during the cooling of the universe, right after decoupling, the beta decay processes, involving the electron mass, change the composition of matter and electrons picks up thermal mass from the hot fermion loop. At T>4 MeV, the renormalization scheme of perturbative QED may not be a very good theory as beta decay contributes through weak interactions, which may cause hard thermal loops and the associated singularities in QED. Therefore, all of our discussions in the following sections are referred to below decoupling temperature.

## Second Order Contribution

Second order thermal corrections to the electron mass come from the two-loop diagrams. The overlapping diagrams give overlapping hot terms with divergent cold terms and the calculations become really cumbersome. However, in the limiting cases, simpler expressions can be obtained, both for low T and high T limits. In the limit T<m, the second order thermal contribution to electron mass [9,10] is

$$\frac{\delta m^2}{m} \simeq (\frac{\delta m^1}{m})^2 + \frac{2\alpha^2 T^2}{3m^2}\left[\frac{33}{2} + \frac{1}{v}\left(\frac{8m}{E} - 1\right) + \left(\frac{5}{v} - \frac{1}{2} - \frac{4m}{Ev^2}\right)\ln\frac{1+v}{1-v}\right.$$
$$\left. - \left\{1 + \frac{1}{2}\left(1 + \frac{1}{v}\right)\ln\frac{1+v}{1-v}\left[\gamma - \ln 2 + \frac{6}{\pi^2}\sum_{r=1}^{\infty}\frac{1}{r^2}\ln\left(\frac{rm}{T}\right)\right]\right\}\right]. \quad (14)$$

This equation shows that the low temperature expression for thermal contribution is very complicated, due to the overlapping hot and cold terms at the two-loop level, as compared to the thermal one loop contribution which is just $\alpha \pi T^2/m^2$. However, the leading order low temperature second order contribution is simply

$$\frac{\delta m^2}{m} \simeq (\frac{\alpha \pi T^2}{3m^2})^2 + \frac{10\alpha^2 T^2}{3m^2} \quad (15)$$

whereas the first term indicates the contribution from the disconnected graph which is usually expected from the iteration method. The second term in this expression is clearly dominant for T<m. In the limit T>m, the electron mass contribution is a long expression and can be found in Ref.[9,10] in detail. Numerical evaluation is not simple and we postpone it for now. However, the leading order contribution at T>m can be written as

$$\left(\frac{\delta m^{(1)}}{m}\right)^2 \simeq \alpha^2\left[\mathcal{M}_1\left(\frac{T}{m}\right)^4 + \mathcal{M}_2\left(\frac{T}{m}\right)^3 + \mathcal{M}_3\left(\frac{T}{m}\right)^2 + \mathcal{M}_4\left(\frac{T}{m}\right) + \mathcal{M}_5\right]. \quad (16)$$

The coefficients $\mathcal{M}$'s in eqn. (16) are complicated functions of electron mass, energy and velocity of electron giving the leading order contribution as

$$\left(\frac{\delta m^{(1)}}{m}\right)^2 \simeq \alpha^2\left(\frac{T}{m}\right)^5 \quad (17)$$

The second order contribution in the above equations is just a leading order contribution to prove that the second order contribution cannot be higher than the first order contribution below the decoupling temperature only. Higher order terms can blow up even at the lower temperatures.

## Wavefunction Renormalization

The electron wavefunction in QED\ is related to the selfmass of electron through the Ward identity. The factor (1-A) is required for renormalization, because then the propagator can also be renormalized by replacing

$$\frac{1}{\not{p} - m + i\varepsilon} \rightarrow \frac{Z_2^{-1}}{\not{p} - m + i\varepsilon}.$$

Thus, for Lorentz invariant self-energy, the wavefunction renormalization constant can equivalently be expressed as

$$Z_2^{-1} = 1 - A = 1 - \frac{\partial \Sigma(p)}{\partial \not{p}}. \tag{18}$$

The fermion wavefunction renormalization in the finite temperature field theory can be obtained in a similar way as discussed in vacuum theories. However, the Lorentz invariance in the finite temperature theory is imposed by setting A=B in eq. (5). Thus using eqs. (14) and (5), one obtains [8]

$$Z_2^{-1}(m\beta) = Z_2^{-1}(T=0) - \frac{2\alpha}{\pi} \int_0^\infty \frac{dk}{k} n_B(k) - \frac{3\alpha}{\pi} b(m\beta)$$
$$+ \frac{\alpha T^2}{\pi v E^2} \ln\frac{1+v}{1-v} \left\{ \frac{\pi^2}{6} + m\beta a(m\beta) - c(m\beta) \right\}. \tag{19}$$

giving the low temperature values as

$$Z_2^{-1} = Z_2^{-1}(T=0) - \frac{2\alpha}{\pi} \int \frac{dk}{k} n_B(k) + \frac{\alpha \pi T^2}{6E^2} \frac{1}{v} \ln\frac{1-v}{1+v} \tag{20}$$

and high temperature value as

$$Z_2^{-1} = Z_2^{-1}(T=0) - \frac{2\alpha}{\pi} \int \frac{dk}{k} n_B(k) + \frac{\alpha \pi T^2}{4E^2} \frac{1}{v} \ln\frac{1-v}{1+v} \tag{21}$$

For small values of the electron velocity $v$, thermal contributions to the wavefunction renormalization constant can be determined from Eqs. (20) and (21) as

$$Z_2^{-1} = Z_2^{-1}(T=0) - \frac{2\alpha}{\pi} \int \frac{dk}{k} n_B(k) - \frac{\alpha \pi T^2}{3E^2} \tag{22}$$

for low temperature, and

$$Z_2^{-1} = Z_2^{-1}(T=0) - \frac{2\alpha}{\pi} \int \frac{dk}{k} n_B(k) - \frac{\alpha \pi T^2}{2E^2} \tag{23}$$

for high temperature.

The finite part of eqs (20) and (21) is equal to δ m/m at E=10m in the relevant temperature range. These terms are suppressed at large values of electron energy E and they are suppressed by a factor $T^2/m^2$. However, the calculated value at that temperature is significantly different. The difference in the thermal contribution can easily be found to be about 50 % of the low temperature value and around 33 % of the high temperature value, just as in δ m/m. This difference can be mentioned as

$$\Delta(Z_2^{-1}) \approx \frac{\alpha \pi T^2}{6E^2} = 3.8 \times 10^{-3} \frac{T^2}{E^2} \qquad (24)$$

The finite part of the wavefunction renormalization constant can be obtained by finding a ratio of temperature with the Lorentz energy E. The minimum value of this energy is equal to mass. Following eq. (13), the higher order contributions to the wavefunction can then be written as

$$Z_2^{-1} = Z_2^{-1}(T=0) - \frac{2\alpha}{\pi} \int \frac{dk}{k} n_B(k) + \exp(\frac{\alpha \pi T^2}{2E^2}). \qquad (25)$$

Eq. (25) indicates that thermal contributions to the wavefunction are always smaller than δ m/m because E is always greater than m. At T = m, and even at the temperatures higher than nucleosynthesis, convergence of the series can be established as T < E at those temperatures. So the two interesting physical limits give smaller thermal contribution in electron wavefunction as the relevant temperature limits can be defined as m<T<E and T<m<E, which ensures the renormalizeability of QED at comparatively higher temperature as compared to selfmass.

### Second order thermal corrections to the electron wavefunction

The renormalization of the wavefunction is directly related to the selfmass of electrons and the guaranteed finiteness of electron mass at finite temperatures, below the decoupling temperature, ensures the finiteness of the wavefunction automatically. However, the detailed expression for the wavefunction renormalization constant can be found in Ref. [9] and can be given as

$$Z_2^{-1} \overset{T<m}{\to} 1 + \frac{\alpha}{4\pi}\left(4 - \frac{3}{\varepsilon}\right) - \frac{\alpha}{4\pi^2}\left(I_A - \frac{I^0}{E}\right) - \frac{\alpha^2}{4\pi^2}\left(3 + \frac{1}{\varepsilon}\right)I_A + \frac{2\alpha^2 T^2}{3\pi^2 m^2}, \qquad (26)$$

Similarly, the high temperature limit for the wave function renormalization constant gives:

$$Z_2^{-1} \xrightarrow{T>m} 1 - \alpha\left[\frac{2I_A}{\pi} + \frac{1}{4\pi}\left(\frac{3}{\varepsilon} - 4\right) + \frac{4\pi T^2}{3}\right]$$
$$- \alpha^2\left[\frac{1}{4\pi^3}\left\{\frac{3}{\varepsilon}(I_A + J_A) + (3I_A + 5J_A) - \frac{8T^2}{3m^2}\right\}\right.$$
$$+ \frac{1}{8}\sum_{n,r,s=1}^{\infty}(-1)^r T\left\{e^{-r\beta E}\left[f_+(s,r)\frac{\gamma\cdot\mathbf{p}}{E^2 v^2} - \frac{I_B I_C}{64\pi^2}\right.\right.$$
$$+ h(p,\gamma)\{f_-(n,r)\frac{I_C}{8\pi} - f_-(s,r)\} + f_+(n,r)\frac{\gamma\cdot\mathbf{p}}{E^2 v^2}\frac{I_B}{8\pi}$$
$$+ \left\{\frac{2}{m}h(p,\gamma) - \frac{\gamma\cdot\mathbf{p}}{E^2 v^2}\right\}f_-(s,r)\Big]\Big\}\Big]. \tag{27}$$

in terms of the one loop integrals I's and J's which can be found in the original papers Ref.[9.10].

**Photon Selfenergy and QED Coupling Constant**

The selfenergy of photons and the electron charge also behave differently for a cooling and a heating system around T=m. It is well-known that the electron charge and the coupling constant do not show significant temperature dependence for T<m. However, they have significant thermal contributions at high temperatures (T>m). Differences in the behavior of a cooling and a heating QED system start again near T~m from the lower side and near the decoupling temperature from a high side. This difference in the coupling constant looks more natural due to the beta decay processes during nucleosynthesis.

Calculations of the vacuum polarization tensor show that the real part of the [9,10] longitudinal and transverse components of the polarization tensors can be evaluated, in the limit $\omega \longrightarrow 0$, as:

$$\operatorname{Re}\Pi_L^\beta(k,0) = \frac{4\pi\alpha}{3}\left\{T^2 + \frac{k^2}{2\pi^2}\ln\frac{m}{T} + \ldots\right\}, \tag{28a}$$

$$\operatorname{Re}\Pi_T^\beta(k,0) = \frac{2\alpha}{3\pi}\left\{k^2 \ln\frac{m}{T} + \ldots\right\}. \tag{28b}$$

giving the interaction potential, in the rest frame of the charged particle as [7]

$$V(k) \equiv e_R^2 \delta_{\mu 0}\left[\frac{u_\mu u_\nu}{k^2 - \frac{4\pi\alpha}{3}\left\{T^2 + \frac{k^2}{2\pi^2}\ln\frac{m}{T}\right\}} + \frac{g_{\mu\nu} - u_\mu u_\nu}{k^2 - \frac{2\alpha}{3\pi}k^2\ln\frac{m}{T}}\right], \tag{29}$$

$$V(k) = e_R^2 \delta_{\mu 0}\Delta_{\mu\nu}\delta_{\nu 0},$$

$e_R$ is renormalized charge in vacuum V(k) can be expanded, at low temperature as:

$$V(k) \equiv e_R^2 \left(1 + \frac{2\alpha}{3\pi} \ln \frac{T}{m}\right) \left[\frac{u_0^2}{k^2 + \frac{4\pi\alpha T^2}{3}} + \frac{1-u_0^2}{k^2}\right]. \tag{30}$$

The constant in the longitudinal propagator is the plasma screening mass, therefore, the outside factor corresponds to the charge renormalization and in turn to the coupling constant. We may then write the coupling constant at low temperatures as:

$$\alpha(T) = \alpha(T=0)\left(1 + \frac{2\alpha}{3\pi} \ln \frac{T}{m}\right) = \alpha(T=0)\left(1 + 1.55 \times 10^{-3} \ln \frac{T}{m}\right) \tag{31}$$

The factor $1.55 \times 10^3 \ln(T/m)$ is a slowly varying function of temperature and does not give any significant contribution near the decoupling temperature and remains insignificant for a large range of temperature, due to the absence of significant numbers of hot electrons in the background. Therefore, the coupling constant is not modified at T<m at all. The temperature dependent factor in the longitudinal propagator $(4\alpha\pi T^2/m^2)$ is the plasma screening frequencies or selfmass of photons that contribute to the QED coupling constant at finite temperatures. For generalized temperatures, the charge renormalization constant $Z_3$ can be written as [10]

$$Z_3 = 1 - \frac{2e^2}{\pi^2}\left\{\frac{c(m\beta)}{\beta^2} - \frac{ma(m\beta)}{\beta} - \frac{1}{4}\left(m^2 - \frac{\omega^2}{3}\right)b(m\beta)\right\}. \tag{32}$$

Also, the electric permittivity is

$$\varepsilon(K) \simeq 1 + \frac{4e^2}{\pi^2 K^2}\left(1 - \frac{\omega^2}{k^2}\right)\left\{\left(1 - \frac{\omega}{2k}\ln\frac{\omega+k}{\omega-k}\right)\left(\frac{c(m\beta)}{\beta^2} - \frac{ma(m\beta)}{\beta}\right) \right.$$
$$\left. - \frac{1}{4}\left(2m^2 - \omega^2 + \frac{11k^2 + 37\omega^2}{72}\right)b(m\beta)\right\}, \tag{33}$$

and the magnetic permeability is

$$\frac{1}{\mu(K)} = 1 + \frac{2e^2}{\pi^2 k^2 K^2}\left[\omega^2\left\{1 - \frac{\omega^2}{k^2} - \left(1 + \frac{k^2}{\omega^2}\right)\left(1 - \frac{\omega^2}{k^2}\right)\frac{\omega}{2k}\ln\frac{\omega+k}{\omega-k}\right\}\left(\frac{c(m\beta)}{\beta^2} - \frac{ma(m\beta)}{\beta}\right)\right.$$
$$\left. - \frac{1}{8}\left(6m^2 - \omega^2 + \frac{129\omega^2 - 109k^2}{72}\right)b(m\beta)\right]. \tag{34}$$

In the limit T>m, the wavefunction renormalization constant can be written as:

$$Z_3 = 1 + \frac{\alpha T^2}{6m^2} \tag{35}$$

giving the renormalized coupling constant as

$$\alpha = \frac{e^2/(\hbar c)}{4\pi\epsilon_0}\left(1 + \frac{\alpha T^2}{6m^2}\right) = \frac{\mu_0 e^2 c}{2h}\left(1 + \frac{\alpha T^2}{6m^2}\right) \tag{36}$$

Eq. (35) gives $Z_3 = 1+1.2\times 10^3\, T^2/m^2$ and leaves the perturbation series valid for at least T≤4m. It is clear from eqs. (31-35) that the coupling constant is basically changed due to the hot fermion loop contributions. Hot bosons do not change the coupling constant as the vacuum fluctuations occur due to fermion loops, at the first order in α. However, the situation is different for higher order contributions.

At T > 4m (the decoupling temperature), thermal contributions are significant enough to grow the coupling constant to the level where it can create a problem for the convergence of the perturbative series of QED. In that case, we need to use non-perturbative methods to establish the renormalization of QED at high temperatures. However, thermal corrections to the coupling constant are significant until the temperature is of the order of electron mass. When it is lower than the electron mass and the primordial nucleosynthesis almost stops α attains the constant value (1/137). It happens because the constant thermal contribution from fermion background $(c(m\beta) = -\pi^2/12)$ at T > m become negligible as soon as the universe cools down to T<m.

### Second order correction to the QED coupling constant

The first order thermal corrections do not contribute to the coupling constant at T<m but they do not vanish at the two-loop level because of the overlap of the vacuum term and the thermal term. The low-temperature second order correction to coupling constant can be give as,

$$Z_3 = 1 + \frac{\alpha^2 T^2}{6m^2} \tag{37}$$

and the high temperature contributions are given as

$$\alpha_R = \left\{ \alpha(T=0) + \frac{8\alpha}{\pi m^2}\left[ \frac{ma(m\beta)}{\beta} - \frac{c(m\beta)}{\beta^2} + \frac{b(m\beta)}{4}\left(m^2 + \frac{1}{3}\omega^2\right)\right] \right.$$
$$+ \frac{\alpha^2}{m^2}\Big[\frac{T^2}{6} + \sum_{n,r,s=1}^{\infty} (-1)^{s+r} e^{-s\beta E} \frac{T}{(n+s)}\Big\{\frac{24T}{(r+s)}$$
$$\left. - 12m^2\Big[\frac{e^{-m\beta(s+r)}}{m} - (r+s)\beta\,\text{Ei}\{-m\beta(r+s)\}\Big]\Big\}\Big]\right\}. \tag{38}$$

Since the contribution to the coupling constant is always proportional to $T^2/m^2$ and is sufficiently smaller than α, that ensures the renormalizeability at the temperatures below the decoupling temperature.

In the next section, we will summarize these results and discuss the behavior of QED in thermal background, below the decoupling temperature.

### Results and Discussion

A quantitative study of the QED renormalization constants, at finite temperatures, show that all the renormalization constants are finite at T≤ 4m. Also the higher order radiative corrections are smaller than the lower order perturbative corrections in real-time formalism. In this range of temperature, the largest thermal contributions come from the mass renormalization constant. The wavefunction renormalization constant is suppressed, especially at low temperatures and relativistic energies. There is no low temperature contribution to the electron charge and QED coupling constant, as the hot fermion loop contributions are ignorable at low temperatures (T<m). However, a rapid growth in the renormalization constants at large temperatures indicate increasing thermal corrections with the increase of hot electrons in the background. A comparison between the one-loop and two loop contributions is shown to prove renormalizeability of the theory, including thermal corrections. What happens around T ≥ m, is expressed in terms of a(mβ), b(mβ) and c(mβ) functions, at the first loop level. These functions give vanishing contributions at low temperatures as they arise from the integration of the hot fermion propagator. Contributions of c(mβ) vanish at low T and it sums up to $(-\pi^2/12)$ for large T values. However, at the two loop level, thermal contributions overlap with the vacuum terms and expressed in a much more complicated overlapping series as well as the same a(mβ), b(mβ) and c(mβ) functions. The problem arises when diverging vacuum terms overlap with the temperature dependent terms. However, they are renormalizable below decoupling temperatures. The low temperature and high temperature contributions are derivable from these general expressions.

We plot thermal contributions to electron mass ($\delta m/m$), electron wavefunction renormalization constant $(z_2^{-1} + \frac{2e}{\pi} \int \frac{dk}{k} n_B(k))$ and the electron charge ($Z_3-1$), that can be derived for T<m and T>m ranges from the same eqs, (8), (18) and (32), respectively, at the one loop level. Since we are dealing with the exponential functions in this study, even less than an order of magnitude difference is a safe limit to use low temperature and high temperature limits. Just to prove the renormalizeability, we present the plots of low temperature and high temperature terms of renormalization constants, as they are derived from general expressions. We give a comparison between a heated and a cooled QED system, around the common temperature of the order of electron mass (See Figure1).The difference between thermal background contributions of a heating and cooling system change due to the difference of background during the heating and cooling process. A cooling system such as the early universe starts off with more fermions and keeps losing them during the cooling process, whereas a heating system will start with a minimum number of fermions and they will be created during the heating process. We just quantitatively discuss the selfmass contribution as it changes the physically measurable mass, a very important parameter of the theory.

Figure 1 indicates the difference between the selfmass of electron due to the low and high temperature first order corrections. It is found that the difference between two values is equal to 1/3 of the low temperature value and 1/2 of the high temperature value at T=m. Similarly the electron wavefunction and charge for a system that approaches from lower temperature to T=m will be different from the one reaching to the same temperature by cooling of a hotter system. It is understandable as the photons do not couple with each other directly. They only couple to the charged fermions and the presence of charge fermion is required to affect the QED coupling. The fermion distribution function (eq.(2)) actually brings in the fermion loop contributions at high T when more fermions are generated during the nucleosynthesis and are not ignorable any more in the medium.

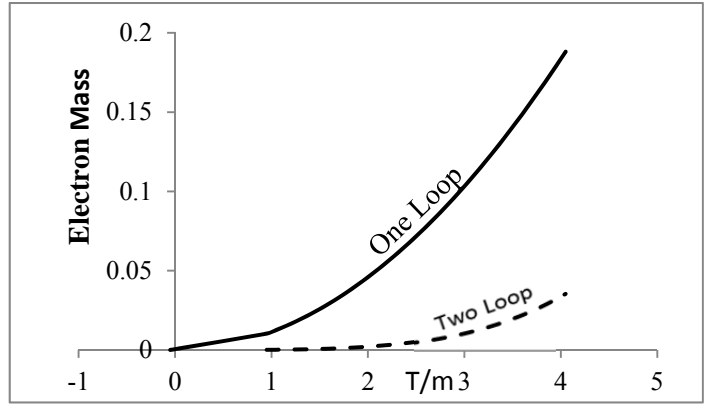

*Figure 2: Comparison of selfmass contribution from the hot background at the one loop (solid line) and at the two loop (broken line) level below the decoupling temperature*

We compare thermal contribution to the first order [9] and the second order [10] corrections to the selfmass of electron in Figure 2. The solid line corresponds to the first order corrections and the broken lines to the second order corrections. Figure 3 gives a similar graph for the renormalized coupling constant of QED in a hot medium. A plot of these renormalization constants shows that the temperature corrections do not affect the renormalizability of the theory for T sufficiently smaller than m up to the two-loop level. The distinction in behavior starts near T~ m without affecting the renormalizability. Due to the small contribution, the behavior of renormalization constants, for both orders, is comparable at low temperatures, indicated by the Figures 2 and 3. This difference is significant at high temperatures. The plot of the QED coupling constant as a function of temperature shows the very small effect at low temperatures, at both loops. It only becomes significant for T>m. In the approximations, used in this paper, one loop thermal contribution is zero at the first loop level at low temperature. However, at high temperatures, the coupling constant grows quadratically with temperature, expressed in units of electron mass. With a large coupling constant, the renormalizability of the theory cannot be guaranteed and non-perturbative methods have to be used to treat the hard thermal loops. However it is not needed under the decoupling temperatures at all.

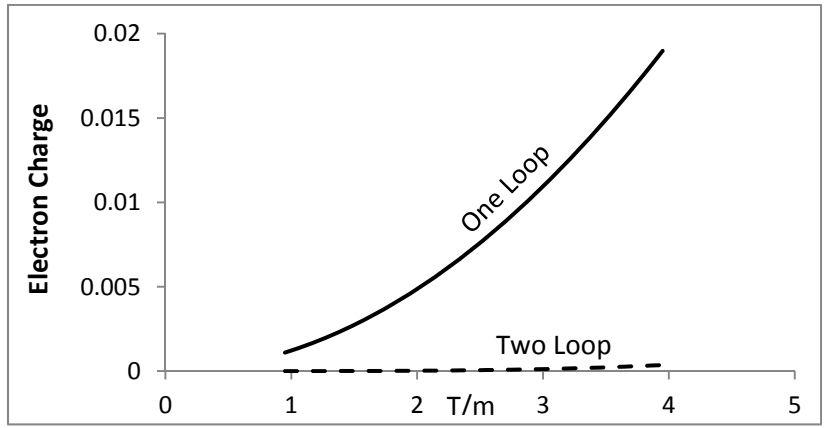

*Figure3: Comparison of QED coupling constant contribution from the hot background at the one loop (solid line) and at the two loop (broken line) level below the decoupling temperature.*

Table 1 shows the leading order thermal contributions to the numerical values of the renormalization constants for electron mass and the QED coupling constant near the decoupling temperatures. It also shows that the thermal contribution of the renormalization constants at temperatures around the decoupling temperature become significant. It is a table of the leading order contributions to prove the renormalizeability. So we consider E≤ 4m, just to get an approximate thermal contribution to electron wavefunction.

*Table 1:The values of the selfmass of electron and QED coupling small constant below the decoupling temperature*

| T/m | Electron Mass | | Charge | Wavefunction for E=10m | |
|---|---|---|---|---|---|
| | *LowT* | *High T* | *High T* | *LowT* | *High T* |
| 0.02 | 3.06E-06 | | | -7.13948E-15 | |
| 0.2 | 0.000306 | | | -7.13948E-11 | |
| 1 | 0.007647 | 0.01147 | 1.6E-07 | -4.46218E-08 | -1.5059E-07 |
| 2 | | 0.04588 | 2.56E-06 | | -2.4094E-06 |
| 3 | | 0.10323 | 1.3E-05 | | -1.2197E-05 |
| 4 | | 0.18352 | 4.1E-05 | | -3.855E-05 |
| 5 | | 0.28675 | 0.0001 | | -9.4116E-05 |
| 6 | | 0.41292 | 0.000207 | | -0.00019516 |
| 7 | | 0.56203 | 0.000384 | | -0.00036156 |
| 8 | | 0.73408 | 0.000656 | | -0.0006168 |
| 9 | | 0.92907 | 0.00105 | | -0.00098799 |
| 10 | | 1.147 | 0.0016 | | -0.00150585 |

However, it is explicitly shown in the above figures and Table1 that the renormalization scheme of QED works below the neutrino decoupling temperature. Above the decoupling temperature, a large number of hot electrons in the background lead to the failure of the QED renormalization scheme at larger temperatures and the electroweak theory has to be incorporated.

The cooling universe of the standard big bang model behaves differently after the neutrino decoupling. Nucleosynthesis starts [17,22] right after the neutrino decoupling and the helium synthesis takes place when the temperature of the universe is cooled down to the temperature of electron mass. This is actually a temperature, where the finite temperature corrections to QED parameters are significant but complicated enough to evaluate it numerically. However, the temperature dependent QED parameters are needed to describe the observations of WMAP data. After the nucleosynthesis is complete, the temperature dependence is back to the quadratic dependence on temperature, though it is not exactly the same as at the low temperature. Fermion background contribution can easily be seen.